\def \SAIT #1 #2 {{\em Mem.\ Soc.\ Astron.\ It.\/} {\bf #1}, #2}
\def \MESS #1 #2 {{\em The Messenger\/} {\bf #1}, #2}
\def \ASTRNACH #1 #2 {{\em Astron. Nach.\/} {\bf #1}, #2}
\def \AAP #1 #2 {{\em Astron. Astrophys.\/} {\bf #1}, #2}
\def \AAL #1 #2 {{\em Astron. Astrophys. Lett.\/} {\bf #1}, L#2}
\def \AAR #1 #2 {{\em Astron. Astrophys. Rev.\/} {\bf #1}, #2}
\def \AAS #1 #2 {{\em Astron. Astrophys. Suppl. Ser.\/} {\bf #1}, #2}
\def \AJ #1 #2 {{\em Astron. J.\/} {\bf #1}, #2}
\def \ANNREV #1 #2 {{\em Ann. Rev. Astron. Astrophys.\/} {\bf #1}, #2}
\def \APJ #1 #2 {{\em Astrophys. J.\/} {\bf #1}, #2}
\def \APJL #1 #2 {{\em Astrophys. J. Lett.\/} {\bf #1}, L#2}
\def \APJS #1 #2 {{\em Astrophys. J. Suppl.\/} {\bf #1}, #2}
\def \APSS #1 #2 {{\em Astrophys. Space Sci.\/} {\bf #1}, #2}
\def \ASR #1 #2 {{\em Adv. Space Res.\/} {\bf #1}, #2}
\def \BAIC #1 #2 {{\em Bull. Astron. Inst. Czechosl.\/} {\bf #1}, #2}
\def \JSQRT #1 #2 {{\em J. Quant. Spectrosc. Radiat. Transfer\/} {\bf #1}, #2}
\def \MN #1 #2 {{\em Mon. Not. R. Astr. Soc.\/} {\bf #1}, #2}
\def \MEM #1 #2 {{\em Mem. R. Astr. Soc.\/} {\bf #1}, #2}
\def \PLR #1 #2 {{\em Phys. Lett. Rev.\/} {\bf #1}, #2}
\def \PASJ #1 #2 {{\em Publ. Astron. Soc. Japan\/} {\bf #1}, #2}
\def \PASP #1 #2 {{\em Publ. Astr. Soc. Pacific\/} {\bf #1}, #2}
\def \NAT #1 #2 {{\em Nature\/} {\bf #1}, #2}

\documentstyle[twoside]{memsait}
\input epsf.sty
\begin{opening}
\title{RXTE Observation of the Seyfert 2 Galaxy NGC4507}
\author{M.Guainazzi$^1$, G.Matt$^2$, L.Piro$^3$, N.R.Robba$^4$}
\institute{$^1$A.S.I.-BeppoSAX Science Data Center, Via Corcolle 19, 00131, Rome, Italy\\
$^2$Dipartimento di Fisica ``E.Amaldi'', Universita' di Roma 3, Via della Vasca Navale, Rome, Italy\\
$^3$I.A.S./C.N.R., Via E. Fermi 21, 00044, Frascati (Rome), Italy\\
$^4$Unita' G.I.F.C.O./C.N.R., Istituto di Fisica, Via Archirafi 36, 90123 Palermo, Italy\\}
\date{} % DO NOT INSERT ANY DATE HERE !!!
\end{opening}

\begin{document}

%\oddpagefooter{\sf Mem. S.A.It., Vol. ??, ??}{}{\thepage}
%\evenpagefooter{\thepage}{}{\sf Mem. S.A.It., Vol. ??, ??}
\oddpagefooter{}{}{} % LEAVE AS IT IS !
\evenpagefooter{}{}{} % LEAVE AS IT IS !
\ 
\bigskip

\begin{abstract}
Preliminary results of the RXTE observation of the Seyfert 2 Galaxy
NGC4507 are presented. The observed broadband [4-100 keV] spectrum is
intrinsically hard ($\Gamma \simeq 1.2$); an iron line
is detected with a relatively high equivalent width ($EW \simeq 400 \ eV$).
The remaining calibration uncertanties are briefly discussed, as well as
the scientific implications of our results.
\end{abstract}

\section{The {\it Bruno Rossi} XTE Observatory}
NGC4507 was observed by {\it RXTE} ({\it Bruno Rossi X-ray Timing Explorer},
Bradt, Rothschild and Swank 1993)
from February 28 to March 10 1996. In the following paper preliminary results
of the data analysis are presented.

The {\sc FTOOLS} 3.5.2 version was used to produce the following results.
Data have been screened according
to intervals of stable pointing (within $\sim 20$ arcseconds from the nominal one)
and elevation angle higher than 10$^{\circ}$.

Proportional Counter Array (PCA)
background spectra estimate procedures are still
in progress; available calibration files could account only
for $\sim 70\%$ of the observed background counts.
Thus, some of the relevant calibration
files were provided by courtesy of PCA hardware group team for weak source
calibration purposes. See {\verb!http://lheawww.gsfc.nasa.gov/users/keith/ngc4507/ngc4507.html!}  for a deeper discussion on the
PCA background strategies and relevant outcomes.
Data only from the first
Xenon layer were collected and [4-60 keV] energy range adopted.

On the other hand, the Crab High Energy X-ray Timing Experiment (HEXTE)
spectrum shows no deviation larger
than 5\% from a simple power law fit
in the energy range [40-250 keV]. However,
a clear $\simeq 10\%$ deviation is visible in the
residuals at lower energy than the Xe
edge ($E \simeq 35 \ keV$).
Moreover the background subtracted NGC4507 HEXTE spectrum
shows an anomalous increase of the counts above 100 keV.
We decided therefore to restrict the HEXTE data analysis
to the energy range [40-100 keV].

\section{Spectral data analysis}

NGC4507 is
the hardest object in the Ginga sample \cite{Smith95}
($\Gamma = 1.39^{+0.70}_{-0.30}$); only very loose upper limits
could be got on the reflected vs. direct component
normalization ($R < 3.0$) and the presence of an iron line
($EW < 990 \ eV$). A more recent ASCA observation \cite{Vignali97}
yielded best-fit values that were nearer to the typical values for
Seyfert 2 spectra ($\Gamma = 1.75\pm0.14$, $EW =
190\pm40 \ eV$), although consistent with the Ginga outcomes within
the statistical uncertainties. The [2-10 keV] flux is in the range
$(1.6-2) \times 10^{-11} \ erg \ cm^{-2} \ s^{-1}$.  

\begin{figure}
\label{fig1}
\epsfysize=8cm % fix the y-dimension and scales x-dim. to y-dim.
%\epsfysize=8cm % fix the x-dimension and scales y-dim. to x-dim.
% Feel free to do the choice you prefer but do not exceed the x-dimension
% of the text lines
\hspace{2.5cm}\epsfbox{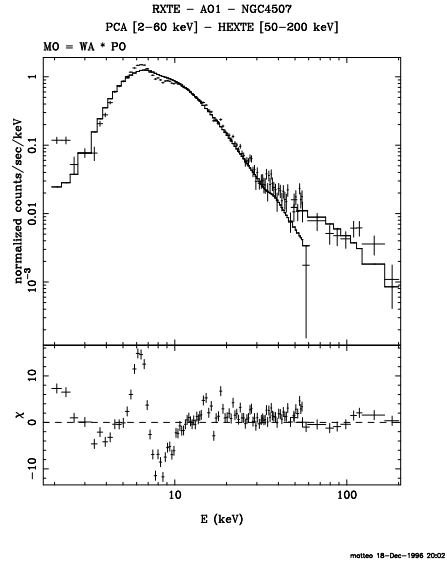} %centering:act on the hspace argument 
\caption[h]{NGC4507 PCA+HEXTE spectrum ({\it upper panel}) and residuals
({\it lower panel}) when a simple power-law model is applied. HEXTE data
points above 100 keV are also displayed to show the suspicious
increase of the flux counts in the background subtracted spectrum,
but were not used in the fits.}
\end{figure}

In Figure~\ref{fig1} the whole braodband PCA+HEXTE spectrum and
the residuals when a simple absorbed power-law model is applied are shown.
The spectrum is
quite hard ($\Gamma = 1.2\pm0.2$) while the 2-10 keV flux is
slighlty lower than the previous measures ($F = 1.2 \times 10^{-11} \
photon \ cm^{-2} \ s^{-1} \ keV^{-1}$). A clear emission-like feature
is visible at energies $E \simeq 6.2 \ keV$, consistent with the 
K${\alpha}$ fluorescence line from neutral matter after taking into account
a residual
$\sim 5\%$ systematic uncertanties in the
gain relation. Such iron line is about twice that expected by emission from
the line--of--sight absorbing material provided that the covering factor
is a significant fraction of 4$\pi$ \cite{Awaki91}, \cite{Ghisellini94},
and would then require a significant iron overabundance.  An iron photoionization edge
is also detected with an optical depth
$\sim 0.4$ and energy consistent with the ionization state that can be
derived from the line centroid energy.

The contemporary presence of hard spectrum, fluorescence emission line
and photoionization edge is also suggestive of a Compton reprocessed continuum,
even if the simple
power law plus the line is formally a satisfactory fit to the data. 
The reflection {\sc XSPEC} model {\sc PEXRAV}
has been used to test such an hypotesis. The results are:
$N_H$=$(34.1^{+0.8}_{-1.4}) \times 10^{22} cm^{-2}$,
$\Gamma_{dir}$=$1.95^{+0.09}_{-0.12}$, $N_{refl}/N_{dir}$=$6.6^{+2.5}_{-1.8}$ 
and $EW_{Fe line}$=$370^{+40}_{-30}$.

The spectral index of the
intrinsic power-law turns out to be consistent with the one typically
observed in Seyfert 1 X-ray spectra \cite{Nandra94}; the Hydrogen
equivalent column density ($N_H \sim 3 \times 10^{23}$) and iron line
EW ($\simeq 370 \ eV$) are consistent with
the Ginga observation outcomes. However
RXTE data requires a huge amount of reflection, the ratio between
the reflected and direct components lying in the range [4.8,9.1].
That could be a result of a delayed response of the reflecting matter to
variation of the continuum (we actually detected the source at a smaller
than usual flux level). The expected value of the iron line for such a value
of $R$ is about twice the observed best--fit value. $R$ and $EW$ are only
marginally consistent with the model expectations.

Alternative models fails to fit the continuum shape.
Neither a partial covering, nor a double absorber produce a satisfactory
fit of the data but for the trivial values of the added parameters.
No exponential cut-offs are
required by the data. If the energy cut-off is fixed at
$E_{cut-off} = 70 \ keV$ (as suggested by \cite{Bassani95}), the required
spectral index is unplausibly high ($\Gamma = 0.8$) and the HEXTE
data points remain sistematically below the model.

\section{Discussion}

The RXTE spectrum of NGC4507 appears rather flat ($\Gamma$=1.2).
Hard spectra have already been observed in other Compton--thin Seyfert 2's
[e.g. NGC5252,
\cite{Cappi96}; NGC7172, \cite{Matt97}; see also Smith \& Done 1996]; even
if for each of these sources alternative explanations to an
intrinsically flat spectrum cannot be ruled out, it is nevertheless
rather embarassing for unification models. For NGC4507, the only 
physically reasonable alternative accepted by the data is the presence
of a huge ($R=6$) Compton reflection component. The large value of $R$ could be
interpreted as delayed response of the reflection component to variations
of the primary continuum, an explanation consistent by the fact that
we have caught the source at a lower flux level than usual, and that
the spectrum during a 1994 ASCA observation had a more typical spectral
index \cite{Vignali97}. The observed iron line is smaller than expected,
suggesting an iron underabundance (it is worth noting that if this line
would be interpreted as due to the line--of--sight absorbing matter, iron
overabundance is required).

\begin{figure}
\label{fig2}
\epsfysize=8cm % fix the y-dimension and scales x-dim. to y-dim.
%\epsfxsize=8cm % fix the x-dimension and scales y-dim. to x-dim.
% Feel free to do the choice you prefer but do not exceed the x-dimension
% of the text lines
\hspace{2.5cm}\epsfbox{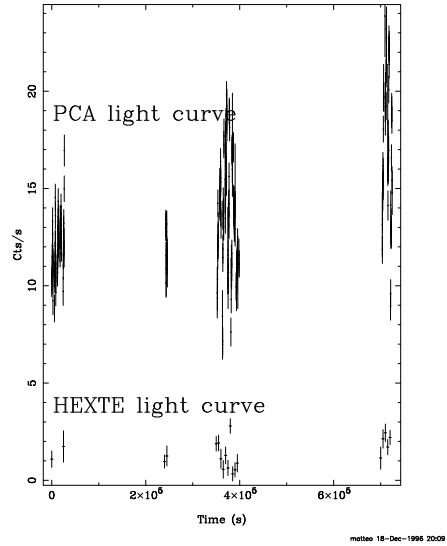} %centering:act on the hspace argument 
\caption[h]{Broadband NGC4507 light curves in the PCA ({\it upper panel})
and HEXTE ({\it lower panel} detectors}
\end{figure}

However, it is necessary
to get rid of any problems concerning the instrumental calibration,
and the
background subtraction procedures (the latter particularly relevant for a 
weak sources like NGC4507) before any final astrophysical conclusion
is drawn. The Figure~\ref{fig2} shows the broadband NGC4507
ligth curve for both the PCA and HEXTE
detectors. Systematics $\sim 30\%$ are still present. Such effects are
likely to be manily due to an uncorrect background subtraction procedure
and are currently still under investigation by RXTE team.

\acknowledgements
The authors warmly thanks Dr. Jahoda for providing still
undelivered calibration files and for many valuable suggestions.

% Example for the references.
%You can use the bibitem tool as specified in the first reference, i.e.
%you can report the reference in the form fixed by the content of the square 
%brackets by inserting in the text of the masnuscript the instruction
%\cite{} with the key defined in the braces as argument. See the exemple.
% Otherwise you can report the references as shown below (2nd to 4th
% references) inserting citations in the text in the standard way.

\end{document}